\documentclass[aps,prl,twocolumn,superscriptaddress,
               amsmath]{revtex4}

\usepackage{graphicx}
\usepackage{dcolumn}
\usepackage{bm}

\begin{document}

\title{Finite Temperature Transitions in Large Magnetic Field in Dipolar Spin Ice}

\author{Jacob P. C. Ruff}
\affiliation{Department of Physics, University of Waterloo, Ontario, N2L 3G1,
Canada
}

\author{Roger G. Melko}
\affiliation{Department of Physics, University of California, Santa
Barbara, California, 93106}

\author{Michel J. P. Gingras}
\affiliation{Department of Physics, University of Waterloo, Ontario, N2L 3G1,
Canada
}
\affiliation{Canadian Institute for Advanced Research, 180 Dundas Street
West, Toronto, Ontario, M5G 1Z8, Canada}

\date{\today}

\begin{abstract}
  We use Monte Carlo simulations to identify the mechanism that allows
  for phase transitions in dipolar spin ice to occur and survive for
  applied magnetic field, ${\bf H}$, much larger in strength than that
  of the spin$-$spin interactions. In the most generic and highest symmetry 
  case, the spins on one out of four sublattices of the pyrochlore decouple 
  from the total local exchange+dipolar+applied field. 
  In the special
  case where ${\bf H}$ is aligned perfectly along the [110] 
  crystallographic direction, spin chains
  perpendicular to ${\bf H}$ show a transition to ${\bf q}=X$ long range
  order, which proceeds via a one to three dimensional crossover. We
  propose that these transitions are relevant to the origin of specific
  heat features observed in powder samples of the Dy$_2$Ti$_2$O$_7$ spin
  ice material for ${\bf H}$ above 1 Tesla.
\end{abstract}

\maketitle

\begin{figure}
\begin{center}
\includegraphics[height=5.5cm]{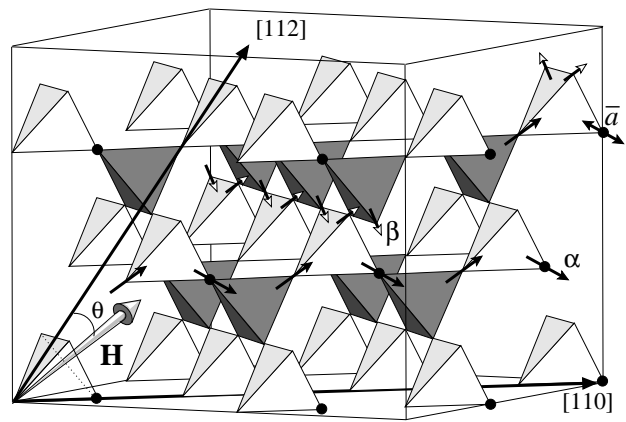}
\caption{Portion of the pyrochlore lattice, with 
the $[110]$  and $[112]$ cubic crystallographic axes.
The double headed arrow shows a $[\bar 1 \bar 1 1]$ spin on a
site of the $\bar a$ sublattice that is decoupled from the
shown ${\bf H}$ (solid arrow).
Sites on the $\bar a$ FCC sublattice (solid black circles) are shown.
}
\label{fig1}
\end{center}
\vspace{-1cm}
\end{figure}

In 1997, Harris {\it et al.} found the surprising experimental result that,
despite overall ferromagnetic interactions of approximately 2 K in strengh, 
the Ho$_2$Ti$_2$O$_7$ magnetic pyrochlore material 
fails to order down to a temperature of 50 mK~\cite{Harris-PRL-1997}. 
In Ho$_2$Ti$_2$O$_7$ the Ho$^{3+}$ magnetic moments reside on a lattice of corner-shared
tetrahedra (Fig.~\ref{fig1}).
It was argued that the strong local Ising anisotropy forcing the Ho$^{3+}$ moments 
to point along the local $\langle 111\rangle$ cubic directions frustrates
the ferromagnetic interactions \cite{Harris-PRL-1997,Harris-JPC}.  
The number of ground states in this frustrated Ising 
system is macroscopic, giving a residual zero temperature 
entropy close to that
estimated by Pauling for the proton disordered state in common hexagonal water ice
I$_h$~\cite{Pauling};
hence the name {\it spin ice} \cite{Bramwell-SCIENCE}.
Specific heat measurements on Dy$_2$Ti$_2$O$_7$ \cite{Ramirez-NATURE},
another Ising pyrochlore material, and Ho$_2$Ti$_2$O$_7$ \cite{Bramwell-PRL,Cornelius}
have provided strong evidence that the low temperature state of these 
systems indeed possess an entropy close to that found by Pauling for water ice.

For water I$_h$, applied hydrostatic pressure introduces constraints that lift the proton 
degeneracy, 
giving a rich pressure-temperature phase diagram and a multitude 
of crystalline ice phases~\cite{crystal-ice}. 
This leads naturally to the question of how to introduce
constraints in magnetic spin ice and to search for analogous ordering behavior. 
Harris {\it et al.}~\cite{Harris-PRL-1997},  
and  Ramirez {\it et al.}~\cite{Ramirez-NATURE} first explored
this question by investigating the behavior of 
Ho$_2$Ti$_2$O$_7$~\cite{Harris-PRL-1997} 
and Dy$_2$Ti$_2$O$_7$~\cite{Ramirez-NATURE} under applied magnetic fields.
In particular, specific heat measurements on powder 
samples of Dy$_2$Ti$_2$O$_7$ revealed three prominent features~\cite{Ramirez-NATURE}. 
For a field ${\bf H}$ larger than 1 Tesla (T),
a sharp peak suggesting a phase transition occurs at a field-independent
temperature of 0.35 K,  surviving up to the largest field 
considered (6 T).  
Another  rounded peak at 1.2 K first appears at 0.75 T and also remains 
observable up to 6 T, although it is less sharp than the
0.35 K feature.  Finally, a third peak at 0.5 K first appears
at 1 T, where it is quite sharp, but disappears for ${\bf H} \gtrsim  $ 3 T.
 Over the past six years, much
theoretical effort has been devoted to the search
for a microscopic explanation of the field-induced spin correlations 
responsible for these specific heat features in powder samples 
of Dy$_2$Ti$_2$O$_7$~\cite{Ramirez-NATURE}.

Building on the work of Harris {\it et al.} on the effect of a $[110]$ 
field  (see Fig.~\ref{fig1}) on Ho$_2$Ti$_2$O$_7$~\cite{Harris-PRL-1997},
Ramirez {\it et al.} suggested that some of these
features might be due to phase transitions 
arising in crystallites that accidently happen to be oriented such that 
some of the Ising spins in the unit cell are perpendicular to ${\bf H}$.
These ``field-decoupled'' spins would then be free to interact among themselves,
giving transitions at temperatures that are field-independent up to a very 
large field~\cite{Ramirez-NATURE}. A similar proposal had been made earlier 
for the
case of garnet systems~\cite{Landau}.
Experimental evidence~\cite{Maeno-111} suggests that the 0.5 K feature 
 in powder samples
arises due to a multicritical point at a temperature of 0.5 K 
and field of 1 T along the $[111]$  crystallographic direction, which is 
related to the interesting phenomenology of kagom\'e ice~\cite{Moessner}.
However, to date, there is no concensus as to whether or not the specific heat
features at 0.35 K and 1.2 K are
caused by the magnetic field directed on crystallites 
of particular orientation~\cite{RMreview,Hiroi,Yoshida}.
In this paper we address this question by using
the {\it dipolar spin ice model}, 
where Ising spins on the pyrochlore lattice interact via
nearest-neighbor exchange and long range magnetic dipole$-$dipole
 interactions~\cite{Hertog}.  We find a generic scenario for field-induced 
phase transitions which survive even when the field 
is much larger than the combined exchange and dipolar interactions.

It is now well established that the nearest-neighbor exchange in
Ho$_2$Ti$_2$O$_7$~\cite{Bramwell-PRL}
and 
Dy$_2$Ti$_2$O$_7$~\cite{Hertog,Fukazawa,Fennell} 
is antiferromagnetic and that the spin-ice phenomenon
originates from long-range 
magnetic dipole-dipole interactions
rather than from nearest-neighbor ferromagnetic exchange
as originally proposed~\cite{Harris-PRL-1997,Harris-JPC}.
The magnetic moment, $\mu_{\rm eff}$,
 of Ho$^{3+}$ and Dy$^{3+}$ in the above materials
is $\mu_{\rm eff} \sim 10 \mu_{\rm B}$, giving a dipolar
coupling constant, $D$, at nearest-neighbor distance of approximately  
1.4 K~\cite{Hertog,Bramwell-PRL}. 
The pyrochlore lattice is a non-Bravais lattice with four atoms per unit cell 
(e.g. a white tetrahedron in Fig.~1)
with each sublattice making a regular FCC lattice.
The 
dipolar spin ice model Hamiltonian is \cite{Hertog}:
\begin{eqnarray}
{\cal H}&=&-\mu_{\rm eff} \sum_{i,a} {\bf S}_{i}^{a} \cdot {\bf H}
\;\; -\sum_{\langle (i,a),(j,b)\rangle}J_{i,a;j,b}{\bf S}_{i}^{a}\cdot{\bf S}_{j}^{b}  \\
&+&  Dr_{{\rm nn}}^{3}\sum_{\tiny \begin{array}{c} i>j \\ a,b \end{array}  }
\frac{{\bf S}_{i}^{a}\cdot{\bf
S}_{j}^{b}}{|{\bf R}_{ij}^{ab}|^{3}} - \frac{3({\bf S}_{i}^{a}\cdot{\bf
R}_{ij}^{ab}) ({\bf S}_{j}^{b}\cdot{\bf R}_{ij}^{ab})}{|{\bf R}_{ij}^{ab}|^{5}}. \nonumber
\end{eqnarray}
The first term is the Zeeman interaction between the spins and the magnetic field ${\bf H}$.
The second and third terms are the exchange and long-range
dipolar interactions, respectively.
The spin vector ${\bf S}_{i}^{a}=\sigma_{i}^{a}\hat{z}^a$
at ${\bf R}_i^a={\bf R}_i+{\bf r}^a$
labels the Ising moment of magnitude
$|{\bf S}_{i}^{a}|=1$ at FCC unit cell coordinate  site ${\bf R}_i$ and
tetrahedral sub-lattice site coordinate ${\bf r}^a$ ($a=1,2,3,4$),
where the local $<111>$ Ising axis is denoted
by ${\hat z}^{a}$ and the Ising variable is $\sigma_{i}^{a} = \pm 1$.
The vector ${\bf R}_{ij}^{ab}={\bf R}_{i}^{b}-{\bf R}_{i}^{a}$ connects spins ${\bf S}_{i}^{a}$
and ${\bf S}_{j}^{b}$.  If one restricts exchange coupling to nearest-neighbor
spins, $J_{i,a;j,b}\equiv J_1$ represents the exchange energy and $D$ the dipolar energy scale.
$J_{i,a;j,b}>0$ is ferromagnetic, $J_{i,a;j,b}<0$ is antiferromagnetic
($J_1>0$ and $D=0$ in the nearest-neighbor spin ice model 
\cite{Harris-PRL-1997,Harris-JPC}).
Previous comparison between Monte Carlo simulations of ${\cal H}$ with
experimental specific heat data has found that
$J_1 \approx -3.72$ K
for Dy$_2$Ti$_2$O$_7$~\cite{Hertog} and $J_1 \approx -1.65$ K 
for Ho$_2$Ti$_2$O$_7$~\cite{Bramwell-PRL}, with $D \approx 1.41$ K for both materials.
In addition, we assume strictly classical spins and neglect quantum 
fluctuations induced by the applied transverse field which admix the 
excited crystal field states with the ground state doublet.  This is a 
reasonable approximation for fields smaller than 10 T, since the 
energy gap to the lowest excited states is $\sim  300$ K in 
Dy$_2$Ti$_2$O$_7$ ~\cite{Rosenkranz}.
\begin{figure}[ht]
\begin{center}
\includegraphics[width=0.45\textwidth]{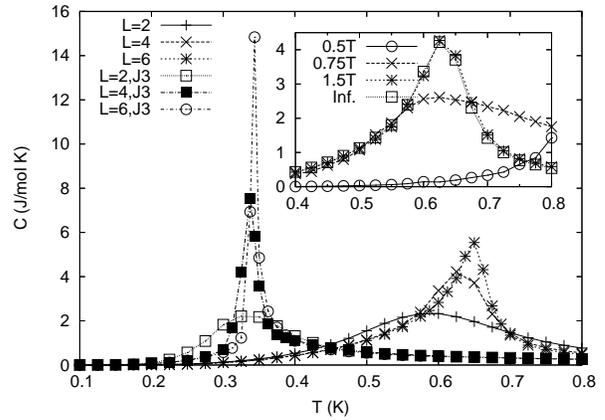}
\caption{
Specific heat from MC simulations with only the ($\bar a$) FCC sublattice
occupied by spins.   
The values $J=-3.72$ K and $D=1.41$ K appropriate for
Dy$_2$Ti$_2$O$_7$ yield a transition at 0.65 K.
Adding a small $J_3 = -0.022$ K shifts this transition
to a temperature 
that agrees with experiments (see text). 
$L$ is the linear size of the system, the
total number of spins in a pyrochlore lattice is
$N=16L^3$.  Inset:  Simulations of the full 
pyrochlore lattice with ${\bf H}(\theta)$ (see text). 
The system behaves like the $\bar a$ FCC lattice in the limit of large (Inf.)
field.}
\label{fig3}
\end{center}
\vspace{-5mm}
\end{figure}

Consider first the problem of spins in the dipolar spin ice model,
subject to an external magnetic field ${\bf H}$,
from the perspective of symmetry.  
For ${\bf H}\ne 0$, the simplest  scenario for a phase transition arises
when one of the four FCC sublattices,  
call  it $\bar {a}$ (see Fig. 1), is decoupled from ${\bf H}$. 
In other words,
the field is perpendicular to the Ising quantization direction
$\hat z^{\bar {a}}$  such that ${\bf H}\cdot \hat z^{\bar {a}} =0$. 
With the condition ${\bf H}\cdot \hat z^{\bar {a}}=0$,
there exists one completely free rotational degree of freedom in the 
orientation of the crystal with respect to ${\bf H}$, which
allows all spins belonging to the $\bar a$ sublattice 
to interact among themselves and participate in a phase transition.  
One obvious field orientation that satisfies this condition is the $[112]$ 
direction, first investigated in Ref.~\cite{Maeno-112}.
Measurements 
on a single crystal of Dy$_2$Ti$_2$O$_7$ with ${\bf H}$ aligned along the $[112]$ 
direction revealed two broad specific heat peaks and no sign of a phase 
transition~\cite{Maeno-112}.
For ${\bf H}$ larger than 0.5 T or so, an anomaly develops at 
$T_{112}^{\rm low} \sim 1.7$ K, independent of field  strength,
while a higher temperature anomaly occurs at $T_{112}^{\rm high} \propto \vert{\bf H} \vert $.
We confirmed the results of Higashinaka {\it et al.} via Monte Carlo simulations 
on the spin ice model for Dy$_2$Ti$_2$O$_7$.
In particular, 
using the
above $J_1$ and $D$ values for  Dy$_2$Ti$_2$O$_7$,
we find that for an infinite $[112]$ field a broad 
Shottky-like specific heat anomaly occurs at $\sim 1.7$ K,
with no sign of a sharp phase transition. 

 We now explain the results in Ref.~\cite{Maeno-112}.
In the limit of strong $[112]$ field,
the spins on the three $a \ne \bar a$ sublattices
that have a spin component parallel to the field are pinned by ${\bf H}$ at 
low temperatures in a ``one-in'' and ``two-out''  spin configuration 
(see top right tetrahedron in Fig.~\ref{fig1}).
Because of the
dipolar and exchange interactions, the spins on the $a\ne \bar a$ sublattices
produce an {\it internal} local  microscopic
field, ${\bf H}_{\bar 1 \bar 1 1}^{\rm mic}$ that enforces the 
two-in$-$two-out ``spin ice rule''~\cite{Bramwell-SCIENCE} acting on the 
spins at sublattice $\bar a$, hence
forcing them to  point ``in''.
 A simple estimate gives the (internal) Zeeman energy scale to be
${\rm H_{\bar 1 \bar 1 1}^{\rm mic}}
 = 2(5D+J_1)/3 \sim 2.22$ K at nearest neighbor distance, modified to 1.71 K using 
the Ewald method to sum dipolar contributions~\cite{RMreview}.
In other words, the candidate would-be field-decoupled $\hat z^{\bar a} = 
[\bar{1}\bar{1}1 ]$ spins, while experiencing
zero applied external field ${\bf H}$, are subject to an inward internal microscopic
field of magnitude
 ${\rm H}_{\bar 1 \bar 1 1}^{\rm mic}=(1.71\,{\rm K})/(7.09\,{\rm K T}^{-1}) = 0.24$ T 
parallel to 
$[\bar 1 \bar 1 1 ]$.

Such a large ${\rm H}_{\bar 1 \bar 1 1}^{\rm mic} \sim 1.7$ K  precludes a
phase transition at $T\ll {\rm H}_{\bar 1 \bar 1 1}^{\rm mic}$.
However, a scenario for a low temperature  phase transition can be reintroduced.
Take the {\it external} applied field ${\bf H}$  canted away from perfect 
alignment perpendicular
to $\hat z^{\bar a} =[\bar{1}\bar{1}1]$ (see Fig.~\ref{fig1}, ${\bf H}$ arrow),
towards the inside of the tetrahedron (i.e. towards the $[110]$ direction), 
at an angle $\theta$ such that
$\theta(\vert{\bf H}\vert) \approx
{\rm H}_{\bar 1 \bar 1 1}^{\rm mic}/{\vert {\bf H} \vert}$ in the limit of 
large applied  field strength.
For this orientation of ${\bf H}$, the net local (internal $+$ external)
field acting on the $\bar a$ sublattice geometrically vanishes,  so that 
 the spins on $\bar a$ become (free) paramagnetic species.
Spins on this $\bar a$ FCC sublattice 
are  actually third nearest-neighbors on the pyrochlore structure 
(see Fig.~1), and therefore
predominantly interact among themselves via long-range dipolar 
interactions (assuming momentarily zero third nearest-neighbor exchange $J_3$).
Interestingly, this field-decoupling scheme freezes out (decimates) 
 spins on the three $a\ne \bar a$ sublattices at large field,
and consequently  generates a very rare physical realization of an 
FCC Ising (dipolar) magnet~\cite{Roser}.

We now determine the ground state for the spins on the $\bar a$ sublattice.
First, recall that dipolar-coupled 
three-component (Heisenberg) spins  have a ferromagnetic ground state on the 
FCC lattice, with 
an energy that is independent of the magnetization 
direction (as for all cubic lattices)~\cite{ferromagnetic-direction}.
Consequently, with the field-decoupled $\bar a$
spins above residing on an FCC sublattice with Ising direction,
$\hat z^{\bar a}= [ \bar 1 \bar 1 1 ]$, 
we expect a phase transition to a ferromagnetic ground state,
with the bulk magnetization along $\pm [ \bar 1 \bar 1 1]$.
Results from Monte Carlo simulations considering
only spins on the $\bar a$ FCC sublattice,
shown in Fig.~\ref{fig3}, confirm that the ordered state is
indeed ferromagnetic below 0.65 K.
The inset of Fig.~\ref{fig3}
shows the results of simulations on the full pyrochlore lattice, with external
 fields ranging from 0.5 T to 1.5 T and  missaligned from $[112]$ with an
angle $\theta(\vert {\bf H}\vert)$, so as to give 
the geometrical 
cancellation above.  There is no discernable difference between 
the results above 1.5 T and the results obtained when considering only
the $\bar a$ sublattice
(labelled ``Inf.''), confirming the simple geometric arguments above.

The transition we observe is at $\sim$ 0.65 K, not 0.35 K as
found in powder specific heat measurements~\cite{Ramirez-NATURE}.  
However, adding a very small antiferromagnetic third 
nearest-neighbor exchange $J_3=-0.022$ K gives a transition to a ferromagnetic ground state 
at a temperature close to the experimental value of 0.35 K. Interestingly, 
an antiferromagnetic $J_3$ of this 
magnitude ($-0.03 {\rm K} <  J_3 < -0.01 {\rm K}$)
appears able to explain the paramagnetic 
zone boundary scattering observed 
in neutron scattering of  Dy$_2$Ti$_2$O$_7$~\cite{Fennell}.
A more antiferromagnetic (negative) $J_3$ changes the ordered
state to a N\'eel state
with ordering wavevector ${\bf q}=(1/2, 1/2, 0)$. 
The present discussion suggests that measurements on a
well oriented single crystal
could allow for a precise determination of a very small $J_3$ value.
A subsequent analysis of
either DC susceptibility \cite{Fukazawa}, specific heat \cite{Hertog} or
elastic neutron scattering \cite{Fennell} could then
help determine the full set of exchange parameters $J_1\! - \! J_2\! -\! J_3$ in
spin ice materials.

 The above calculation of ${\rm H}_{\bar 1 \bar 1 1}^{\rm mic}$ assumed
static spins on the three $a\ne \bar a$ sublattices, and neglected
thermal fluctuations. 
Conversely, the applied ``compensating'' field parallel to $[11\bar 1$] 
from the canted ${\rm H}(\theta)$ is static.
Hence, for any finite large 
 $[112]$ external field plus weak $[11\bar 1]$ compensating field of 
approximately 0.24 T, there is at nonzero temperature
a net ``overcompensating'' outward field along $[11\bar1]$ due to 
the thermal fluctuation (spin flips) of spins on the three 
$a\ne \bar a$  sublattices. 
The global $\langle 111 \rangle$ Ising
symmetry is therefore always explicitely broken.
The $\theta-{\bf H}$ relationship above is akin to the first
order $({\bf H}=0,T<T_c)$ transition line in a conventional Ising ferromagnet.
By correcting the angle $\theta(\vert {\bf H}\vert)$, evaluated above for 
$T=0$, to the
one that would cancel the average $H^{\rm mic}_{\bar 1 \bar 1 1}$ 
at $T=T_c$, a true spontaneous symmetry breaking can then occur at $T_c$.
For $J_3 < -0.03$, the overcompensating field is not conjugate to the 
N\'eel order parameter
and the transition is thermodynamically sharp for 
large applied ${\bf H}$ perpendicular to $[1 1 \bar 1] $.
\begin{figure}
\begin{center}
\includegraphics[width=0.45\textwidth]{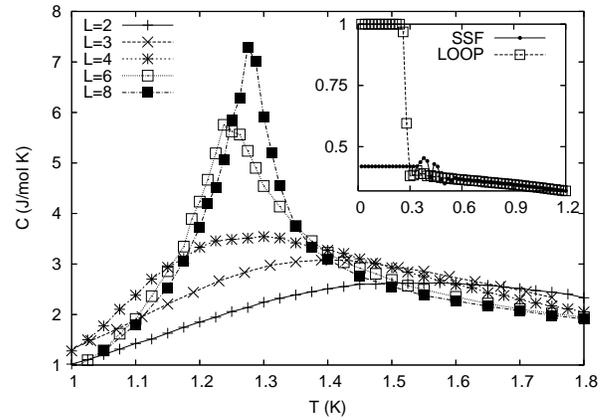}
\caption{
Specific heat from MC simulations of $\beta$ chains, equivalent to the
limit large of $[110]$ ${\bf H}$ applied to a single crystal of Dy$_2$Ti$_2$O$_7$.
The inset shows the ${\bf q}=X$ order parameter obtained via loop
Monte Carlo (loop, open squares)
 \cite{RMreview,Melko} for a $[110]$ field of 0.05 T and
compared with conventional Monte Carlo results using single spin flip
(SSF, filled circle). 
}
\label{fig2}
\end{center}
\vspace{-5mm}
\end{figure}

We now turn to the special case in which a second sublattice
becomes accidentally field-decoupled as the field is rotated in the
plane perpendicular to one of the four $\hat z^{\bar a}$ directions. 
This occurs for ${\bf H}$ along the $[110]$ and related crystallographic 
directions 
\cite{Harris-PRL-1997,Ramirez-NATURE,RMreview,Hiroi,Yoshida,Fennell-field}.
As illustrated in Fig.~\ref{fig1}, the $[110]$ 
field direction couples to the Ising spins on two FCC 
sublattices (forming the $\alpha$ chains), but leaves spins on
the two FCC sublattices forming the $\beta$ chains fully decoupled.
Unlike the generic case above with only one 
($\bar a$) decoupled sublattice, the net local internal field from
static spins on the $\alpha$ chains vanishes on lattice
sites that belong to $\beta$ chains.
Although the spins on the $\beta$ chains are then
in principle free to undergo an ordering transition via some long range
exchange or dipolar interaction,  
experimental~\cite{Harris-PRL-1997,Hiroi,Fennell-field}
and numerical~\cite{Hiroi,Yoshida} evidence for this transition
has so far been elusive.
Our Monte Carlo simulation results on the dipolar spin ice model show 
strong evidence for a transition to the so-called ${\bf q}=X$ 
state~\cite{Harris-PRL-1997} for ${\bf H}$ parallel to $[110]$, where
ferromagnetic one-dimensional ($1D$) $\beta$ chains along $[1 \bar 1 0]$ orient
antiferromagntically with respect to their four nearest neighbour
chains (see Fig.~\ref{fig2}).
This is found by studying both the ${\bf q}=X$ order parameter~\cite{RMreview}
and relative  zero-temperature energies of the various competing 
spin ice ground states~\cite{RMreview,Yoshida}.
A transition to ${\bf q}=X$ occurs for all fields greater than $\sim$ 0.02 T, 
although for weak fields 
(${\bf H}<$ 0.10 T) non-local loop spin flips are needed 
in the simulation to find the 
ground state \cite{Melko}.  For fields weaker than 0.02 T, 
the transition is to the ${\bf q}=(0,0,1)$ 
long-range ordered state found by Melko {\it et al.}~\cite{RMreview,Melko}. 
The critical temperature $T_c^{110}(\vert {\bf H}_{110}\vert)$ for the
 ${\bf q}=X$ ordering is strongly field 
dependent for field strengths between 0.02 T and 0.2 T~\cite{RMreview,Yoshida},
but saturates at a critical temperature of 
$T_c^{110}(\infty) \sim 1.2$  K for ${\bf H} > 0.2$ T, as also recently found 
in Ref. \cite{Yoshida}, but contrary to what was obtained in
Ref. \cite{Hiroi}.
For large ${\bf H}$ along $[110]$ the transition to ${\bf q=}X$ proceeds via a
$1D$ to $3D$ crossover with a correlation length $\xi_{1D}$
of approximately
20 nearest-neighbor lattice spacing at the crossover temperature. 
This crossover at such a large $\xi_{1D}$ makes the study of finite-size 
scaling and 3D critical behavior very demanding at this transition.
The presence of dilute defects or impurities or the lack of perfect 
alignment~\cite{RMreview} of ${\bf H}$
along the $[110]$  direction might be at play in preventing the 
development of perfect ${\bf q}=X$ long range order in 
experiments~\cite{Harris-PRL-1997,Hiroi,Fennell-field}.
Finally, we note that introducing $J_3=-0.022$ K
in Eq.~(1) results in a shift of $T_c^{110}(\infty)$ 
to $T_c^{110}(\infty) \sim 1.6 $ K, in disagreement with 
the position of the specific heat feature at 1.2 K \cite{Ramirez-NATURE,Hiroi}.
  Such a large shift of $T_c^{110}$ is 
consistent with the $T_c^{110}$  calculated using a 
1D chain mean-field theory.
Overall, the results presented in this paper suggest that there might be 
important competition between $J_1$, $J_2$ and $J_3$
exchange in Dy$_2$Ti$_2$O$_7$, and that the original description of 
Dy$_2$Ti$_2$O$_7$ using only a nearest-neighbor exchange 
$J_1=-3.72$ K \cite{Hertog,Fennell} should be revisited.

In conclusion, we have identified a mechanism that
causes
field-independent phase transitions 
in the dipolar spin ice model when subject to a large external applied field.
For polycrystalline Dy$_2$Ti$_2$O$_7$, we suggest
that the transition at $T_c^{[112]} \approx 0.35$ K 
can arise 
from crystallites for which the field is accidentally 
quasi-perpendicular to one of the
$\langle 111 \rangle$ directions, but slightly canted towards the corresponding
$\langle 1 1 \bar 1 \rangle $ direction to cancel internal exchange plus dipolar local field.  
This proposed mechanism could be verified in a single crystal experiment.
We also postulate a small third neighbor antiferromagnetic exchange to account 
for the precise $T_c^{[112]}$.
In addition, we found unequivocal support for
the existence of a transition to a long-range ordered ${\bf q}=X$ state for dipolar spin ice
in a magnetic field  along $[110]$ in excess of 0.02 T, which is postulated to account for
 the $T_c^{[110]} \approx 1.2$K transition in polycrystalline Dy$_2$Ti$_2$O$_7$. 
As an interesting 
 by-product of this work, we have identified 
a novel realization of an Ising FCC dipolar ferromagnet.


We thank S. Bramwell, J. Chalker, M. Enjalran, T. Fennell,
Y. Maeno, R. Higashinaka and
T. Yavorsk'ii for useful discussions.
R.  M. acknowledges financial support from the National Science Foundation, Grant No.
DMR02-11166. M. G. acknowledges support from
the NSERC of Canada, the Canada Research Chair Program, 
Research Corporation and the Province of Ontario.

\end{document}